# Unpacking Blockchains

JOHN PRPIĆ - Faculty of Business Administration, Technology and Social Sciences, Lulea University of Technology

---

## 1: INTRODUCTION

The Bitcoin digital currency appeared in 2009. Since this time, researchers and practitioners have looked "under the hood" of the open source Bitcoin currency, and discovered that Bitcoin's "Blockchain" software architecture is useful for non-monetary purposes too. By coalescing the research and practice on Blockchains, this work begins to unpack Blockchains as a general phenomenon, therein, arguing that all Blockchain phenomena can be conceived as being comprised of transaction platforms and digital ledgers, and illustrating where public key encryption plays a differential role in facilitating these features of Blockchains

## 2: CRYPTOCURRENCY

In 2008, Satoshi Nakamoto[1] released a whitepaper detailing the principles behind a "peer-to-peer electronic cash system" (Nakamoto 2008) called Bitcoin. Early in 2009, Nakamoto followed up the whitepaper with a release of the very first source code for Bitcoin at Sourceforge[2] (Swan 2015), thus ensuring that Bitcoin would be open source software (Lakhani & Von Hippel 2003). While outlining the principles of Bitcoin, Nakamoto notes that prior to Bitcoin, electronic payment systems were always dependent on trusted third parties (i.e. financial intermediaries such as Banks or PayPal) to verify that digital transactions were not duplicated, in what is characterized as the "double-spend" problem of electronic payment (Nakamoto 2008, Anceaume et al 2016, Teutsch, Jain & Saxena 2016, Yli-Huumo et al 2016). The Bitcoin system is then put forward as a solution to the double-spend problem, and in turn, to trusted third party intermediation, by offering a solution based on cryptographic signatures instead of trust (Nakamoto 2008, Luu et al 2016). To date, at least 250 other cryptocurrencies (known as "alt-coins") have been developed, many of which are based upon the open source Bitcoin software (Luu et al 2016). The top nine "alt-coins", when combined with Bitcoin, represent approximately $19 Billion USD in market capitalization[3].

### 2.1 CRYPTOGRAPHIC PRIVACY & AUTHENTICATION

Hash algorithms (Zobrist 1970, Knuth 1973), or more specifically, one-way hash/cryptographic hash functions (Merkle 1989b, Goldreich 2009), apply a mathematical function to data, to convert data of arbitrary size, into a new digital string of a predefined and fixed length - a hash. A hash should be easy to compute in one direction (i.e. from data to hash), but the reverse computation (i.e. from hash to data) should be as difficult as possible (Merkle 1989b). Some well-known cryptographic hash functions in use include MD5, SHA 256[4], SHA1, and SHA2 (Radack 2009, Mazonka 2016), and are used as part of public key cryptography systems (Diffie & Hellman 1976, Simmons 1979, Merkle 1980, ELGamal 1985, Merkle 1989a, Yli-Huumo et al 2016).

Public key or asymmetric cryptography systems (Simmons 1979, Fujisaki & Okamaoto 1999) work by issuing a unique pair of keys to everyone on the system; a public key and a private key. In turn, implementing these two different keys in specific ways, in interaction with the public and private keys of other users, can affect two different outcomes; privacy and authentication (Diffie & Hellman 1979). For privacy, a sender encrypts a message with the public key of the intended recipient, and then makes the encrypted message public (or sends it directly to the intended recipient), who then decrypts the message with their private key. For authentication, a sender encrypts a message with their private key, and then makes this encrypted message public. At this point, anyone else can use the sender's public key to verify that encrypted data was indeed created by the sender. This act of using a private key to encrypt a message, and then making the encrypted message public, for the scrutiny of others (by your public key), is called a digital signature (Diffie & Hellman 1979). Thus, altogether, public key cryptography systems separate secrecy (privacy) and authentication (digital signatures) through these two-different means (Simmons 1979).

---

[1] The origins of Bitcoin are attributed to Satoshi Nakamoto. No individual to date has confirmed that Satoshi Nakamoto is not a pseudonym, and there is much speculation about the actual identity of the Bitcoin founder.
[2] https://sourceforge.net/projects/bitcoin/files/
[3] http://coinmarketcap.com/
[4] https://en.bitcoin.it/wiki/SHA-256

**3: BLOCKCHAINS EMERGE**

In the Bitcoin system (as illustrated in Figure 1 below), Nakamoto defined an electronic coin as "…a chain of digital signatures…" (Nakamoto 2008), that are joined together in certain sized bunches known as blocks, where these blocks are created through the verification of each signature by third parties known as miners. Miners supply the computational power to perform the cryptographic proof of authenticity (or proofs of work) of a transaction, and earn Bitcoins from the Bitcoin system software for completing a block (DuPont 2014). Once a block is acknowledged as complete by the Bitcoin system, the system rewards the miner with the promised number of Bitcoins, and the new block is appended to the very end of the historical chain of blocks, which contains every other block that has ever been completed in the system, in order. This overall historical chain, containing every transaction ever created and verified in the system, has come to be known as a Blockchain (Dziembowski 2015, Forte, Romano & Schmid 2015, De Filippi & Hassan 2016, Glaser 2017).

**Figure 1 - A Chain of Digital Signatures** (from Nakamoto 2008)

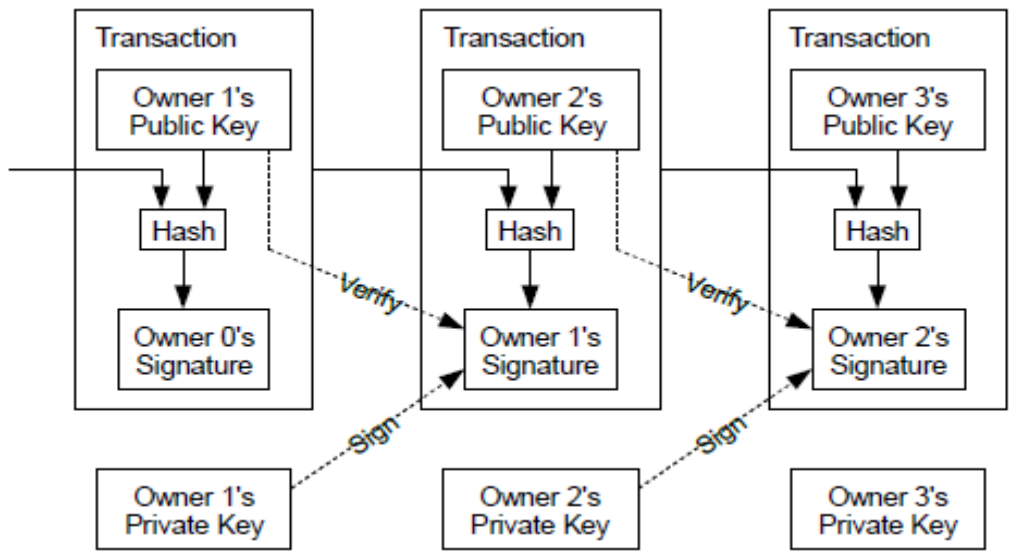

The Bitcoin system is considered to be the very first Blockchain, and has been described as "…a very specialised version of a cryptographically secure, transaction-based state machine" (Wood 2014). In general though, there are many different Blockchains that can be constructed using the same, or similar, processes, and overall, a Blockchain can be thought of as an emergent chain of transactions, where new transactions are validated (by miners), and then inserted to the end of the existing chain of blocks (Forte, Romano & Schmid 2015). In general, a Blockchain process is comprised of the following steps:

1) Collect new transactions and organize them into blocks,
2) Cryptographically verify each transaction in the block,
3) Append the new block to the end of the existing Blockchain.

Other than creating transactions, which is done by users of a Blockchain system, all the above steps are performed by Blockchain miners, through client software that enforces protocol execution using cryptographic mechanisms (Forte, Romano & Schmid 2015). These Blockchain miners have been described as a decentralized network of people controlling computational nodes that validate transactions, and earn cryptocurrency for their efforts (Reijers, O'Brolcháin & Haynes 2016).

**4: TWO ASPECTS OF BLOCKCHAINS: TRANSACTIONS & LEDGERS**

In a Blockchain, every transaction is time-stamped, verified, added in sequence, and made public; from the very first transaction, until the current moment. In turn, every entity interacting (Muftic 2016) with a Blockchain (i.e. miners, senders, receivers, Blockchain software), maintains, and can access, a copy of the entire history of a Blockchain, while knowing that said history cannot be altered in any way, except through new transactions (Reijers, O'Brolcháin & Haynes 2016).

Thus, a Blockchain system not only allows entities to transact securely and directly with one another through public key cryptography privacy; a Blockchain system also creates an immutable, publicly-shared, publicly collected, and publicly verified (and verifiable) record of said transactions in the process, through public key cryptography signatures. This latter aspect of Blockchains has come to be known as a Digital Ledger, and is beginning to be recognized as a valuable (Dziembowski 2015, Reijers, O'Brolcháin & Haynes 2016), and generalizable (Wood 2014) outcome, of all Blockchain systems.

Outside of what is presented in this work, Bitcoin or Blockchain oriented research does not explicitly acknowledge these two separate, but connected, aspects of Blockchain systems, and thus, the research generally focusses on one of these sides or the other. Broadly speaking then, we see literature that has begun to emerge on the transaction-side of Blockchains, focusing on the different types of data that can be transmitted through transactions (Dziembowski 2015, Kraft 2016), transaction signatures (Noether & Mackenzie 2016), the trustless nature of transactions (Wood & Steiner 2016), the privacy of transactions (Kosba et al 2016), the scalability of transactions (Vukolić 2015), the technical processes of transactions (Tschorsch & Scheuermann 2015), market volatility (Ortisi 2016), and incentives to verify transactions (Luu et al 2016). Similarly, we see literature that has emerged on the Ledger-side of Blockchains, focusing on smart contracts (Szabo 1997, Atzei, Bartoletti & Cimoli 2016, Idelberger et al 2016, Piasecki 2016, Reijers, O'Brolcháin & Haynes 2016), Blockchains for Crowdfunding (Jacynycz et al 2016, Zhu & Zhou 2016), Ledger generation rate (Kiayias & Panagiotakos 2015), Ledger temporality (Swan 2016), private Ledgers (Gramoli 2016), Ledgers as consensus systems (Crosby et al 2016), as governance and regulatory technologies (De Filippi & Hassan 2016, Reijers, O'Brolcháin & Haynes 2016), and the economic and legal aspects of Ledgers (Evans 2014, Lee 2016, Schroeder 2016, Sorrell 2016, Zou, Wang & Orgun 2016).

Hence, the recognition that Blockchains simultaneously supply a transaction platform, and an immutable and verified historical record of said transactions, both driven by different aspects of public key cryptography, is a fundamental contribution of this work, and has been illustrated in Figure 2 below.

**5: CONCLUSION**

The aim of this work is to unpack Blockchain phenomena. The work does so, by first investigating the origins of Blockchains (stemming from the Bitcoin system), and outlines that the Bitcoin system solves the double-spend problem of electronic payment, by incorporating public key encryption, to affect both the privacy and authentication of transactions. From here, the work illustrates that the Bitcoin system implements these cryptographic mechanisms to allow transactions to be chained to one another (without a trusted 3[rd] party intermediary), to in turn, create an immutable and verifiable historical record of transaction blocks in the system, that has come to be known as a Digital Ledger. From this point of departure, the work argues that all Blockchain phenomena can be conceived as being comprised of transaction platforms and digital ledgers, and illustrates where public key encryption plays a differential role in facilitating these features of Blockchains.

Altogether, this work provides technical and conceptual clarity for the emerging research on Blockchains. The hope is that future research can leverage this clarity to consistently and effectively investigate Blockchain phenomena in a fine-grained manner.

**Figure 2 - Two Aspects of Blockchains**

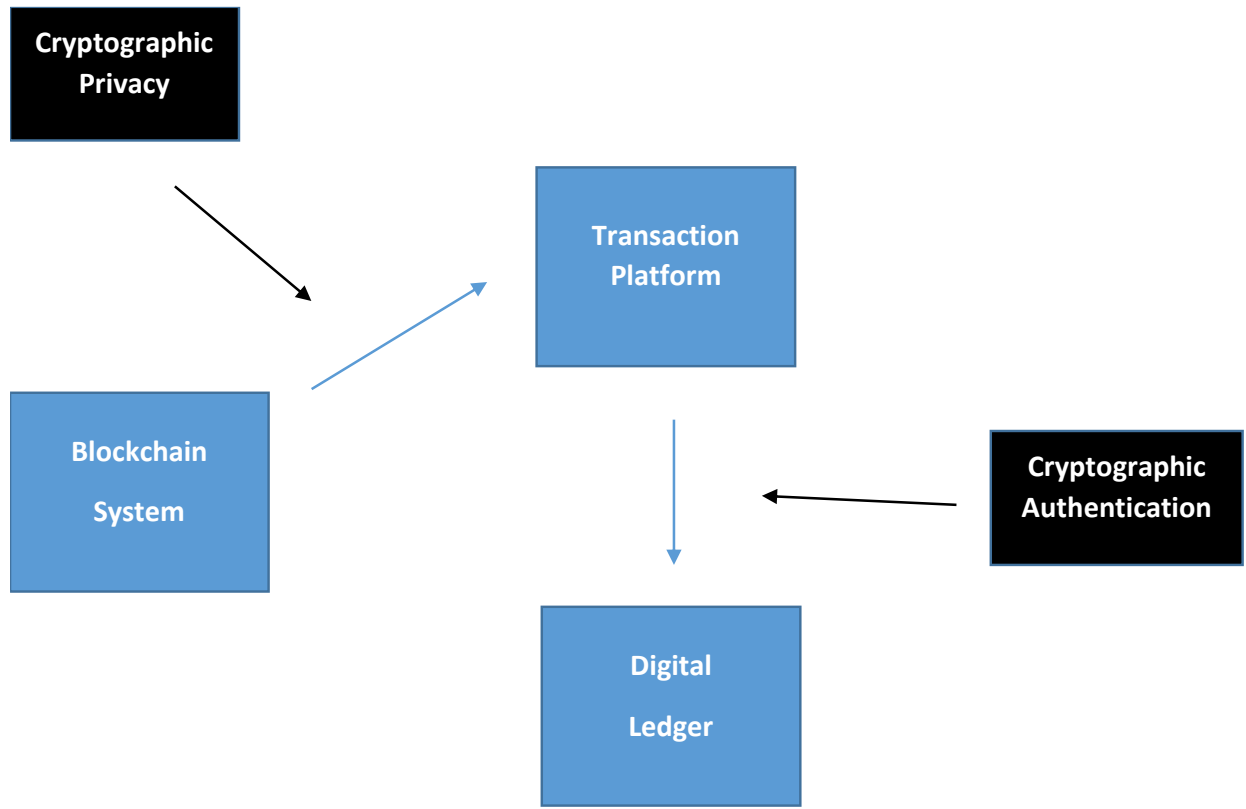

## 6: REFERENCES